\documentclass[aps,prb,twocolumn,showpacs]{revtex4}
\usepackage{graphicx}
\begin{document}

\title{Phase diagram of a surface superconductor in parallel magnetic field}

\author{Ol'ga V. Dimitrova and M. V. Feigel'man}

\affiliation{L. D. Landau Institute for Theoretical Physics, Moscow, 119334, 
Russia}

\begin{abstract}
Universal phase diagram of 2D surface superconductor with generic Rashba 
interaction in a parallel magnetic field is found. In addition to uniform 
BCS state we find two inhomogeneous superconductive states, the stripe 
phase with $\Delta({\bf r})\propto \cos({\bf Qr})$ at high magnetic fields,
and a new "helical" phase with $\Delta({\bf r}) \propto \exp(i{\bf Qr})$ 
which intervenes between BCS state and stripe phase at intermediate 
magnetic field and temperature.
% Component of superfluid density $n_s$  vanishes on the BCS-helical 
%transition line.  
We prove that the ground state for helical phase carries no current.
\end{abstract}


\maketitle

Many efforts both theoretical and experimental have been spent in search 
of exotic nonhomogeneous superconducting states beginning with a pioneering 
papers by Larkin-Ovchinnikov and Fulde-Ferrel (LOFF)~\cite{LO,FF} predicting 
the stripe state in superconductors with competing ferromagnet interaction. 
Nevertheless no convincing experimental evidence for existence of such a state 
was found till now, partially due to rather narrow existence
range of the LOFF state. Recently Barzykin and 
Gor'kov~\cite{BG} did find a system where such inhomogeneous superconducting phase 
could be prominent. It is a two-dimensional surface superconductor with spin-orbital 
Rashba interaction.  One of possible realization of such a system probably was  
reported in very interesting experiments~\cite{NaW}, where various signatures of
surface superconductivity with $T_c \approx 90 K$
were detected in the insulating $WO_3$  doped by small amount of $Na$.
% All three components of this model: kinetic energy, Rashba term 
% and phonon exchange, firmly belong to two-dimensional surface physics, which give 
% confidence that such a system will be eventually found experimentally.
Surface spin-orbit superconductivity is unusual one due to the absence of inversion 
symmetry: this results in the  presence of the the spin-orbital
Rashba term~\cite{rashba} and the chiral subband splitting of free 
electron spectrum at the surface. In such a superconductor
condensate wave-function is a mixture of both singlet and triplet 
states~\cite{Ed,GR}, therefore Pauli susceptibility is not 
vanishing~\cite{GR} at $ T \to 0$;  paramagnetic breakdown of superconductivity 
in a parallel magnetic field is shifted 
towards much higher field values due to the formation of LOFF state~\cite{BG}.
The line of transition from normal to (any of) superconductive state $T_c(h)$ was 
determined in~\cite{BG}; however, the nature of the phase intervening between uniform 
BCS and stripe LOFF state was not studied. This question is important because 
phenomenological theory~\cite{Agterberg} predicts the possibility for a new helical 
state distinct from both BCS and stripe LOFF states.

In this Letter we provide a detailed phase diagram of a surface
superconductor in a parallel magnetic field $h$. 
The phase diagram turns out to be universal after normalization of the temperature and 
the Zeeman energy by the critical temperature in zero magnetic field 
$T_{c0}$ for two models of high (I) and low (II) electron density.
The model I assumes a normal 2D metal with Fermi energy being much greater than chiral
splitting. The model II is suited for an electron gas 
in field effect heterostructures where electrons fill only 
the bottom of one of chiral bands.
We demonstrate the existence of a helical state with 
order parameter $\Delta \propto \exp(i{\bf Qr})$ \, (where 
${\bf Q} \perp {\bf h}$)\, and $Q \sim \mu_B h/v_F$ in a considerable part of
the phase diagram, which  is summarized in Fig.1. 
The line ${\cal L}{\cal T}$ is the second-order transition line separating
helical state from the homogeneous superconductor.
Below the ${\cal T}$ point first-order transition between 
BCS and inhomogeneous state takes place. 
The line ${\cal ST'}$ is the line 
of soft instability of the helical state (see below).
The point ${\cal S}$ in the $T_c(h)$ line is
special in that here the order parameter symmetry is $U(2)$ instead of 
usual $U(1)$.
% Supercurrent response to vector potential component 
% $A_y = {\bf A}{\bf Q}/Q$ vanishes on the ${\cal LT}$ line.
Full details of our theory will be presented in a separate publication~\cite{long}.

Near the surface of a crystal the translational symmetry is reduced and the
inversion symmetry is broken even if it is present in the bulk. As a result a 
transverse electric field appears at interface and gives rise to
the relativistic spin-orbit interaction known as the Rashba term: 
$H_{so}=\alpha\left[\vec{\sigma} \times \hat{\vec{p}}\right]\cdot\vec{n}$,
where $\alpha>0$ is the spin-orbit coupling constant, $\vec{n}$ is a unit
vector perpendicular to the surface, 
$\vec{\sigma}=(\sigma^x,\sigma^y,\sigma^z)$ are spin Pauli matrices.
Spin operator does not commute with the Rashba term, thus 
spin projection is not a good quantum number. On the other hand the 
chirality operator: $\sigma^x\sin\varphi_{\bf p}-\sigma^y\cos\varphi_{\bf p}$, 
does commute with the Hamiltonian, where 
\,$\varphi_{\bf p}$ is the angle between the momentum of the
electron and the $x$-axis. Its eigenvalue:
$\lambda=\pm 1$ is a quantum number of an electron state 
$(\vec{p},\lambda)$. The Rashba term preserves the Kramers degeneracy with states 
$(\vec{p},\lambda)$ and $(-\vec{p},\lambda)$ belonging to the same energy.

In this Letter we consider the simplest model of surface superconductor: 
Gor'kov model for two-dimensional metal with the Rashba term being 
included~\cite{GR}, in the limit $\alpha p_F\gg T_c$. The Hamiltonian
written in the coordinate representation reads
\begin{eqnarray}\label{koord}
\hat{H}=\int \psi_{\alpha}^+(\vec{r})\left ( \frac{\hat{P}^2}
{2m}\delta_{\alpha\beta}
+\alpha\left [ \vec{\sigma}_{\alpha\beta}\times \hat{P}\right ]\cdot \vec{n}- 
\right. \nonumber\\    \left.
-g\mu_B\vec{h}\cdot \vec{\sigma}_{\alpha\beta}/2\right )
\psi_{\beta}(\vec{r})d^2r -\frac{U}{2}\int
\psi_{\alpha}^+\psi_{\beta}^+\psi_{\beta}\psi_{\alpha}d^2r,
\end{eqnarray}
where $m$ is the electron mass and $\alpha$, $\beta$ are spin indices, 
$\hat{P}=-i\vec{\nabla}+\frac{e}{c}\vec{A}(\vec{r})$ is the momentum 
operator in the presence of infinitesimal in-plane vector-potential 
$\vec{A}=\vec{A}(\vec{r})$.
Zeeman interaction with a uniform external magnetic field $\vec{h}$ parallel 
to the interface and in the $x$-direction is included. 
The vector-potential of such a field can be chosen to have only $z$-component, 
therefore it decouples from the 2D kinetic energy term. $\mu_B$ is Bohr magneton
and $g$ is the Lande factor. Hereafter we use a notation $H=g\mu_Bh/2$.
 
The electron operator can be expanded in the basis of plane waves 
$\hat{\psi}_\alpha(\vec{r})=\sum_{\lambda\vec{p}}\ 
e^{i\vec{p}\vec{r}}\hat{c}_{\alpha{\bf p}}$. 
The one-particle part of the Hamiltonian (\ref{koord}) in the momentum 
representation:
\begin{eqnarray}\label{momrep}
\hat{H}_0=\sum_{\vec{p}}\hat{c}_{\alpha {\bf p}}^{+}
\left (\frac{p^2}{2m}+\alpha\left[
\vec{\sigma}_{\alpha\beta} \times \vec{p}\right]\cdot\vec{n}-\vec{H}\cdot
\vec{\sigma}_{\alpha\beta} \right )\hat{c}_{\beta {\bf p}}
\end{eqnarray}
can be diagonalized by the transformation
$\hat{c}_{\alpha{\bf p}}=\eta_{\lambda\alpha}(\vec{p})\hat{a}_{\lambda{\bf p}}$ 
with the two-component spinor 
$\eta_{\lambda}(\vec{p})=(1, i\lambda\exp(i\varphi_{\bf p}))/\sqrt{2}$. 
Eigenvalues of the Hamiltonian (\ref{momrep}) corresponding to the chiralities 
$\lambda=\pm 1$ are
\begin{equation}\label{eigenv}
\epsilon_\lambda(\vec{p})=p^2/2m -\lambda\sqrt{\alpha^2 p^2-2\alpha
p_{y}H+H^2}.
\end{equation}
In the case of high electron density and respectively chemical potential
$\mu\gg m\alpha^2$ ( model I) both chirality branches are filled and the 
equal momentum electron states are split by $2\alpha p_F$. Fermi circles 
with different chiralities are split with the radii 
$p_F= \sqrt{2m\mu+ m^2\alpha^2} \pm m\alpha$. 
Densities of states on the two Fermi circles are almost the same, 
$\nu_{\pm}=\frac{m}{2\pi}\left(1\pm \frac{\alpha}{v_F}\right)$, 
and in this paper we neglect the difference $\nu_+-\nu_-$. 
In the case of the low electron density (model II): 
$-m\alpha^2/2<\mu<0$, the electrons 
fill the bottom of only one chiral branch $\lambda=1$, i.e. a ring. 
The two Fermi circles of the ring in zero magnetic field have radii 
$p_F=m\alpha+l\sqrt{2m\mu+ m^2\alpha^2},$
where $l=\pm 1$ is the number specifying the inner and the outer part of the 
ring and is analogous to chirality. Density of states on the outer and inner
part of the Fermi ring ($l=\pm 1$) is almost the same 
$\nu_{\pm}=\frac{m}{2\pi}\frac{\alpha}{v_F}$ in the case of
narrow ring: $v_F\ll \alpha.$
If magnetic field is applied, the two Fermi circles are displaced in 
opposite $y$-directions by a momentum $Q=\pm H/v_F$, where 
$v_F=\sqrt{2\mu/m+ \alpha^2}$ is the Fermi velocity in the I model, or half
width of the ring divided by $m$ in the II model. 
The pairing interaction (the last term in Eq.(\ref{koord})) can be factorized in 
the chiral basis: 
$\hat{H}_{int}= -\frac{U}{4}\sum_{\bf q}\hat{A}^+(\vec{q})\hat{A}(\vec{q})$, 
where the pair annihilation operator $\hat{A}(\vec{q})=
\sum_{{\bf p}\lambda}\lambda e^{i\varphi_{\bf p}} \hat{a}_{\lambda -{\bf p}_-}
\hat{a}_{\lambda {\bf p}_+}$ with ${\bf p}_{\pm}={\bf p}\pm{\bf q}/2$.

To calculate thermodynamic potential $\Omega=-T\ln{Z}$, we employ
imaginary-time functional integration technique with Grassmanian 
electron fields $a_{\lambda, {\bf p}}$, $\bar{a}_{\lambda, {\bf p}}$
and introduce auxiliary complex field $\Delta({\bf r},\tau)$ to 
decouple pairing term $H_{int}$, cf.~\cite{Popov}. The resulting 
effective Lagrangian is of the form:
\begin{eqnarray}
&&L[a,\bar{a},\Delta,\Delta^*]=\sum_{\bf p}\bar{a}_{\lambda{\bf p}}\left
(-\partial_{\tau}-\epsilon_\lambda(\vec{p})\right )a_{\lambda{\bf p}}+ 
\nonumber \\ &&+\sum_{\bf q} 
\Big[-\frac{|\Delta_{\bf q}|^2}{U}+\frac{1}{2}\sum_{{\bf p}\lambda}
\big(\Delta_{\bf q}\lambda e^{-i\varphi_{\bf p}}\bar{a}_{\lambda,{\bf p}_+}
 \bar{a}_{\lambda,-{\bf p}_-} + \nonumber \\ 
 &&+\Delta^*_{\bf q}\lambda e^{i\varphi_{\bf p}} a_{\lambda,-{\bf p}_-}
 a_{\lambda, {\bf p}_+}\Big].
\label{Lhub-str}
\end{eqnarray}
Below we will work within the mean-field approximation which is 
controlled by the small Ginzburg number ${\rm Gi} \sim T_c/E_F$.
It is equivalent to the saddle-point approximation for the functional 
integral over $\Delta$ and $\Delta^*$, thus we will be studying minima 
of the functional $\Omega[\Delta,\Delta^*]$ appearing after Gaussian 
integration over Grassmanian fields.

\begin{figure}
\includegraphics[angle=0,width=0.477\textwidth]{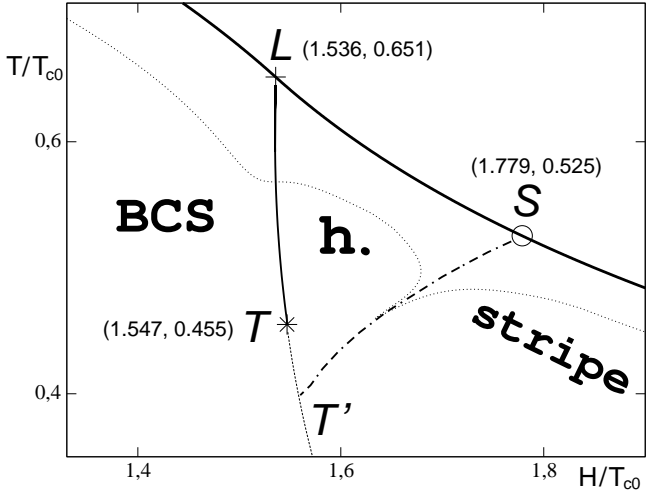}
\caption{\label{phd} Phase diagram that shows:
superconducting phase transition line $T_c(H)$ (bold solid)
and  two second order phase transition lines in the clean case: 
$\cal{L}\cal{T}$ line between homogeneous (BCS) and helical (h.) state and 
$\cal{S}\cal{T'}$ line of stability of helical state.  
Short-dashed line going downwards from the point
${\cal T}$ marks the absolute limit of stability of the BCS state.             
Dotted line shows the physical $T_{BKT}(H)$ line for values
$T_{c0}/\epsilon_F=0.02$ and $\alpha/v_F=0.34$.}
\end{figure}

Near the normal-superconducting phase transition 
the order parameter $\Delta({\bf r})$ is small and
$\Omega$ may be expanded in powers of $\Delta({\bf r})$
and its gradients. We consider the order parameter as a superposition
of the finite number of harmonics, $\Delta({\bf r})= \sum_{i}
\Delta_{{\bf Q}_i}({\bf r})\exp\left(i{\bf Q}_i{\bf r}\right)$,
where $\Delta_{{\bf Q}_i}({\bf r})$ are slowly varying in 2D 
space functions,
and derive the corresponding Ginzburg-Landau functional:
\begin{eqnarray}\label{Ginzburg-Landau}
&\Omega&=\int\Big [\sum_i\alpha_i|\Delta_{{\bf
Q}_i}|^2+c_i\left|\left(-i\frac{\partial}{\partial
\vec{r}}+\frac{2e}{c}\vec{A}\right)\Delta_{{\bf Q}_i}\right|^2 \nonumber \\
&+&\sum_{ijk}\beta_{ijk}\Delta_{{\bf Q}_i}\Delta^*_{{\bf Q}_j}
\Delta_{{\bf Q}_k}\Delta^*_{{\bf Q}_i-{\bf Q}_j+{\bf Q}_k}\Big ]\ d^2\vec{r}
\end{eqnarray}
The coefficient $\alpha(\vec{Q})$ includes Cooper loop diagram
with transferred momentum $Q$:
\begin{equation}\label{C}
\alpha(\vec{Q})=\frac{1}{U}-\frac{T}{2}\sum_{\omega,\lambda,\vec{p}}
 G_\lambda^n \left(\omega,\vec{p_+} \right) G_\lambda^n
\left(-\omega,-\vec{p_-}\right)
\end{equation}
where in the I model the normal-state Green function
in in-plane magnetic field $H\ll\alpha p_F$ is
$G_\lambda^n\left(\omega,\vec{p}\right)= (i\omega-\xi-\lambda
H\sin{\varphi_{\bf p}})^{-1}$,
and $\xi=p^2/2m-\lambda\alpha p_F-\mu$ 
is assumed to be small compared to $\alpha p_F$, whereas in the II model
$H\ll m\alpha^2$ and
$\xi_l=v_F(l\pi-mv_F)$ 
is assumed to be small compared to $m\alpha^2$
($\pi$ is related to the center of the ring). 
Hereafter we present results for high
density (I model), but results for low density are identical.
The condition $\min_{Q}\alpha(\vec{Q})=0$ determines the second-order 
transition line (if $\beta>0$) between the normal metal and the 
superconductor:
\begin{equation}\label{CchT}
\frac{1}{U}=\max_{Q} \sum_{\lambda,\omega>0} \frac{\pi \nu T}
{\sqrt{\omega^2+H_\lambda^2}},
\end{equation}
hereafter $H_\lambda=\lambda H+v_F Q/2.$
Depending upon $H$, the maximum in Eq.(\ref{CchT}) is attained
either at $Q=0$ or at nonzero $\pm |Q|$.
$T_c(H)$ line is shown on Fig.\ref{phd} where both
$T_c$ and $H$ are normalized by the critical 
temperature at zero magnetic field  $T_{c0}=
 2\omega_D\exp(-1/\nu U+C)/\pi$, where $C=0.577$ is the Euler constant. 
Near ${\cal L}$ point $v_F Q(H)\sim \sqrt{H^2-H_L^2}$. 
In the limit $H/T_{c0} \to\infty$ we recover the asymptotics of $T_c(H)$
line \cite{GR}, and find 
$v_F Q(H)=2H-\frac{\pi^4 T_{c0}^4}{7\zeta(3)e^{2C}H^3}$.
Note that $Q=2H/v_F$ is the momentum splitting of two 
$\lambda=\pm 1$ Fermi surfaces in parallel magnetic field.
Lifshitz point ${\cal L}$ separates $Q=0$ and
$Q\neq 0$ solutions on the $T_c(H)$ line; it is the end point of the second
order phase transition between the two superconducting phases. 
At $H > H_L$ inhomogeneous superconductive phase is formed below 
$T_c(H)$ line. No more than 
two harmonics contribute to $\Delta({\bf r})$ just below $T_c(H)$ line:
$\Delta(y) = \Delta_+e^{iQy} + \Delta_- e^{-iQy}$. Below $T_c(H)$
density of the  thermodynamic potential $\Omega$ is lower 
in the superconductive state than in the normal one  by the amount:
\begin{eqnarray}\label{FF}
\Omega_{sn} = \alpha|\Delta|^2 +\beta_{s}|\Delta|^4 + 
\beta_{a}(|\Delta_+|^2 -|\Delta_-|^2)^2,
\end{eqnarray}
where $|\Delta|^2 =|\Delta_+|^2 + |\Delta_-|^2 $. We find coefficients 
$\beta_{s,a}$ using standard diagram expansion around normal state.
At the symmetric point ${\cal S}$ where $\beta_{a}(T_c(H),H)=0$ the
free energy (\ref{FF}) is invariant 
under $U(2)$ rotations of the order parameter spinor $(\Delta_+,\Delta_-)$.
At $ H < H_S$ we find $\beta_{a} < 0$ and the free energy at $T < T_c(H)$
is minimized by the choice of either $\Delta_+ = 0$ or  $\Delta_- = 0$, 
both correspond to helical state. At $H > H_S$ we find $\beta_{a} > 0$ 
and in the free energy minimum $|\Delta_+| = |\Delta_-|$, which corresponds to 
the LOFF-like stripe phase with $\Delta(y) \propto \cos(Qy)$.
At lower temperatures in this phase $\Delta(y)$ contains higher harmonics and 
is time reversal symmetric.
In the helical state with only one harmonics $\Delta(y)=\Delta e^{iQy}$,
the thermodynamic potential reads:
\begin{eqnarray}
\Omega_{hel}= - \nu T\sum_{\omega,\lambda} \int\sqrt{(\omega+
iH_\lambda\sin\varphi)^2+\Delta^2} \frac{d\varphi}{2} + \frac{\Delta^2}{U}. 
\label{SCenergy}
\end{eqnarray}
In equilibrium the stationary conditions 
$\partial\Omega_{hel}/\partial \Delta=0$ and 
$\partial \Omega_{hel}/ \partial Q=0$ are satisfied and they can be found 
explicitly:
\begin{eqnarray}\label{secline}
\frac{1}{2\nu U}=T\sum_{\omega>0, \lambda}\frac{{\bf K} 
\left(k\right)}{r(H_\lambda,\omega)}; \,\quad
\sum_{\omega>0,\lambda}f(H_\lambda,\omega) = 0
%\label{2cond}
\end{eqnarray}
where $r(H_\lambda,\omega)=\sqrt{\omega^2+(|H_\lambda|+\Delta)^2}$, 
the Jacoby modulus $k=2\sqrt{\Delta |H_\lambda|} /r(H_\lambda,\omega)$
and the function $f(H_\lambda,\omega)$ is defined through the Jacoby 
complete elliptic integrals of the first and the second kind,
\begin{equation} f(H_\lambda)=\frac{1}{H_\lambda} 
\left((\omega^2+H_\lambda^2+\Delta^2)\frac{{\bf K}(k)}{r(H_\lambda,\omega)}-
r(H_\lambda,\omega){\bf E}(k)\right).
\label{f}
\end{equation}

We prove by direct microscopic calculation that
$\vec{j}_s=\frac{2e}{\hbar}\partial \Omega/\partial \vec{Q},$ therefore
the equilibrium state carries no supercurrent.
 
Minimum of the thermodynamic potential (\ref{SCenergy})
over $\Delta$ can be expanded in series of small $Q$:
$\Omega_{hel}(Q)=\Omega_{hel}(0)+
\frac{\hbar}{8m}n_s^{yy}Q^2+bQ^4+cQ^6, \quad c > 0.
\label{Omegaexp}$
The condition $n_s^{yy}=0$, $b>0$ determines the second order Lifshitz 
transition line ${\cal LT}$, 
which ends at the point $\cal T$, where coefficient $b=0$ 
changes sign. We compute the
coordinates of $\cal T$ point using Eqs.(\ref{SCenergy},\ref{secline}). 
At lower temperatures $b < 0$ and first-order transition out of BCS 
state occurs.

The domain of helical state local stability was determined via 
consideration of the thermodynamic potential variations due to 
weak static modulation of the form 
$\delta \Delta(\vec{r})= v_{-q}\exp(-iqy) + v_{q+2Q}\exp(i(q+2Q)y)$ 
(the presence of two Fourier harmonics in the
perturbation is due to inhomogeneity of the background helical state):
$\delta\Omega_{\delta v} = \vec{v}^+\hat{\cal A}(q) \vec{v}$,
where $\vec{v}=(\delta v_{-q},\delta v^*_{q+2Q})$ and
\begin{equation}
\hat{\cal A}(q)=\frac{\hat{1}}{U}-\sum_{\omega>0,\lambda,{\bf p}}
\left( \begin{array}{cc} G_{\lambda p_-}
G_{\lambda -p_+} & F_{\lambda p_-} F_{\lambda -p_+} \\ 
F^*_{\lambda p_-} F^*_{\lambda-p_+} &
G_{\lambda p_+ +Q} G_{\lambda-p_- +Q} \end{array}\right).
\end{equation}
The matrix ${\cal A}$ has two eigenvalues $\epsilon_1(q)<\epsilon_2(q)$. 
We define the helical state metastability line ${\cal ST'}$ as a collection 
of points where one mode $\delta v$ becomes energetically favorable: 
${\rm min}_q\epsilon_1(q)=0$. 
We solve numerically four equations simultaneously:
two gap equations (\ref{secline}) that determine equilibrium $\Delta$ and
$Q$, together with the two equations $\partial_q\epsilon_1(q)=0$ and 
$\epsilon_1(q)= 0$. By means of expansion of the Ginzburg-Landau functional 
up to the terms of order $|\Delta_\pm|^8$, we checked that the next-order 
"anisotropic" term in the expansion (\ref{FF}) is of the form 
$\varepsilon (|\Delta_+|^2 -|\Delta_-|^2 )^4$, with $\varepsilon > 0$.
This fact ensures that the phase transition out of the helical state
is of the second order (at least, near the $T_c(H)$ line).

We calculated electromagnetic response function 
$\delta j_\alpha/\delta A_\beta = -\frac{e^2}{m c}n_s^{\alpha\beta}$
for helical state using standard diagram methods and found that
$n_s^{yy} = 4\frac{m}{\hbar}\frac{\partial^2\Omega}{\partial Q^2}$.
Thus on the Lifshitz line $\cal{L}\cal{T}$ 
there is no linear supercurrent in the direction perpendicular to the
magnetic field. The component $n_s^{xx}$ does not vanish anywhere
in helical state region and is of the order of $n_s$ of the BCS state. 
This is in contrast with the classical LOFF problem, where $n_s$ was 
shown to vanish in the whole helical state; the difference is probably 
due to the fact that in our problem the direction of ${\bf Q}$ is fixed 
by an applied field ${\bf h}$, while for the case of ferromagnetic 
superconductor it is arbitrary.
The obtained behaviour of $n_s^{\alpha\beta}$ tensor indicates
strongly anisotropic electromagnetic response of surface superconductor
near the Lifshitz line $\cal{L}\cal{T}$.

So far we discussed the clean case; below we demonstrate that sufficiently
high concentration of nonmagnetic impurities suppress both helical 
and stripe states. Consider impurities with  weak short-range 
potential, characterized by to elastic scattering time $\tau$.
%The interaction between the electrons and the impurity atoms at positions 
%$\vec{R}_i$ corresponds to the Hamiltonian
%\begin{equation}
%\hat {H}_{int}=\sum_{i}\int \psi_{\alpha}^+(\vec{r}) \psi_{\alpha}
%(\vec{r}) u(\vec{r}-\vec{R}_i) d^2r,
%\end{equation}
%where $u(\vec{r}-\vec{R}_i)= u\delta(\vec{r}-\vec{R}_i)$; the elastic 
%scattering time is $\tau=1/2\pi n_{imp}u^2\nu,$
%and $n_{imp}$ is the number of impurity atoms per unit volume.
By means of diagram technique~\cite{AGD}, we calculate the coefficient 
$\alpha(\vec{Q})$ in the Ginzburg-Landau 
expansion (\ref{Ginzburg-Landau}) which is the electron-electron vertex 
in the Cooper channel in the presence of nonmagnetic impurities.
 It is given by a sum of ladder 
diagrams which are an alternating sequence of blocks of two normal metal Green 
functions 
$G_{\lambda}=(i\omega-\xi-\lambda
H\sin{\varphi_{\bf p}}+\frac{i}{2\tau}{\rm sgn}{\omega})^{-1}$
and an impurity line. In every block momenta on the upper
and lower lines are opposite whereas the chiralities are the same. 
$T_c(H)$ line is found from $\min_{Q}\alpha(\vec{Q})=0$, where
$\alpha(\vec{Q})=1/U-\pi \nu T\max_{Q} \sum_{\omega>0}
K\left(\omega,H,\tau,Q\right),$ with the Cooper kernel $K$ given by
\begin{equation}\label{impTc}
K\left(\omega,H,\tau,Q\right)=4\tau \frac{I^0_s\left[1-I^2_s\right]+
(I^1_a)^2}{\left(1-I^0_s\right)
\left[1-I^2_s\right]-(I^1_a)^2},
\end{equation}
where $I^{\gamma}_{s,a}=I^{\gamma}_+\pm I^{\gamma}_-$ ($\gamma=0, 1, 2$)
are functions of $(T, H, \tau, Q)$:
$I_{\lambda}^0=
\left({\tilde{\omega}}^2+H_{\lambda}^2\right)^{-1/2}/4\tau,$
$I_{\lambda}^1=
i\left(\tilde{\omega}I_{\lambda}^0-\frac{1}{4\tau}\right)/H_{\lambda},$
$I_{\lambda}^2=i\tilde{\omega}I_{\lambda}^1/H_{\lambda}$, 
$\tilde{\omega}=\omega+\frac{1}{2\tau}$.
At $H=0$ time-reversal symmetry is recovered and Eq.(\ref{impTc})
simplifies to $K=2/\omega$ independently on disorder, 
in agreement with Anderson theorem. We evaluate numerically 
$\alpha^{''} = \partial^2\alpha(\vec{Q})/{\partial {\vec Q}}^2$
along the the transition line $\alpha(T,H,\vec{Q})=0$.
At $\tau T_{c0}/\hbar \leq 0.11$ we found $\alpha'' >0$ at any $H$,
i.e. both stripe and helical state disappear from the phase diagram at
$\tau \leq \hbar/9T_{c0}$. Similar condition for usual LOFF state
is  more stringent, cf.~\cite{Aslamazov}.
In the dirty limit $\tau T_{c0}\ll 1$ the kernel $K$
simplifies to
$K=2/(\omega+2H^2\tau+v_F^2Q^2\tau/4)$. From this form of $K$ one
easily concludes that the paramagnetic critical field {\it grows} 
with increase of disorder as 
$H_{p}=\sqrt{\pi T_{c0}/4\tau e^C}$, cf. similar 
result in~\cite{Klemm}.

The phase diagram Fig.\ref{phd} was obtained 
neglecting small parity-breaking term of the order 
$\alpha/v_F \ll 1$ in the thermodynamic potential (\ref{SCenergy}):
$\delta \Omega=\eta Q$ with $\eta=-\alpha\nu T\sum_{\omega}f(H,\omega)$,
where function $f$ is defined for the clean case in Eq.(\ref{f}).
Taking this term into account while minimizing $\Omega$, one finds that
uniform BCS state transforms into a weakly helical state 
(predicted phenomenologically in~\cite{Agterberg})
with small wave-vector  $\tilde{Q}\approx\frac{2\alpha H}{v_F^2}$
and without supercurrent (homogeneous state would carry supercurrent, 
as was found in~\cite{Yip}, but it is not the ground-state).
The line of 2nd order transition $\cal{LT}$ broadens then into a sharp 
cross-over region between two helical states with small and large
values of $Q$. Within this crossover region,  the superfluid 
density tensor is strongly anisotropic, with
$n^{yy}_s/n^{xx}_s \sim (\alpha/v_F)^{2/3}$. 
% Minimizing the Cooper kernel $\partial K/\partial Q=0$ shows that 
In the dirty limit long-wavelength helical modulation 
 is present everywhere in  superconducting state; near the transition
line its wave-vector $\tilde{Q} = \frac{4\alpha H}{v_F^2}$.

We have calculated $T_c(H)$ line within mean-field approximation (MFA),
those accuracy is usually of order $T_c/\epsilon_F$ for clean 2D 
superconductor: actual transition is of Berezinsky-Kosterlitz-Thouless
vortex depairing type, and is shifted downwards in temperature by
about $T_c^2/\epsilon_F$.
In our system fluctuations are enhanced strongly
around ${\cal L}$ and ${\cal S}$ points. Near the ${\cal L}$ point it is
due to smallness of $n^{yy}_s$, and the enhancement factor is of the 
order of $\sqrt{n^{xx}_s/n^{yy}_s} \sim (\alpha/v_F)^{-1/3}$. 
In the vicinity of the point ${\cal S}$ fluctuations are enhanced
due to extended $U(2)$ symmetry of the order parameter; 2D renormalization
group calculation shows that $U(2)$ fluctuation modes shift actual $T_c$ 
 by  $\Delta T_c \sim 4(T_c^2/\epsilon_F)\log{\beta_s/\beta_a}$ downwards
at $\beta_a \ll \beta_s$.
In result, phase transition line $T_c(H)$ is deformed in the vicinities
of ${\cal L}$ and ${\cal S}$ points, as shown on Fig.1.

In conclusions, we have demonstrated an existence of three
different superconductive states (two helical states and
stripe state) in clean surface superconductor in
parallel magnetic field, and have located transition lines between them.
Strong disorder eliminates short-wavelength helical and stripe states, 
whereas long-wavelength helical state survives.
We thank P. M. Ostrovsky and M. A. Skvortsov for many useful discussions. 
This research was supported by SCOPES grant, RFBR grant 01-02-17759, 
Russian ministry of science and Program "Quantum macrophysics" 
of RAS. O. V. Dimitrova is grateful to Dynasty Foundation for 
financial support.

\end{document}